\documentstyle[prl,aps]{revtex}\twocolumn
\newcommand{\nc}{\newcommand}
\nc{\cntr}[1]{\begin{center} #1 \end{center}}
\nc{\be}{\begin{equation}}
\nc{\ee}{\end{equation}}
\nc{\ba}{\begin{array}}
\nc{\ea}{\end{array}}
\nc{\bea}{\begin{eqnarray}}
\nc{\eea}{\end{eqnarray}}
\nc{\bml}{\begin{mathletters}}
\nc{\eml}{\end{mathletters}}
\nc{\bfi}{\begin{figure}}
\nc{\efi}{\end{figure}}
\nc{\nn}{\nonumber}
\nc{\pp}{{\prime\prime}}
\nc{\eg}{, $e.\  g.$,~}
\nc{\egb}{$e.\ g.$~}
\nc{\ie}{, {\em i.~e.},~}
\nc{\bi}{\bibitem}
\begin{document}

\draft %Remove % to print pacs numbers!
\title{Parisi Phase in a Neuron}
\author{G. Gy\"orgyi and P. Reimann \protect{\cite{pr}}}
\address{Institute for Theoretical Physics, 
E\"otv\"os University, Puskin u.\  5-7, H-1088 Budapest, Hungary}
\date{\today}
\maketitle
\begin{abstract}

Pattern storage by a single neuron is revisited.  Generalizing
Parisi's framework for spin glasses we obtain a variational free
energy functional for the neuron.  The solution is demonstrated at
high temperature and large relative number of examples, where several
phases are identified by thermodynamical stability analysis, two of
them exhibiting spontaneous full replica symmetry breaking.  We give
analytically the curved segments of the order parameter function and
in representative cases compute the free energy, the storage error,
and the entropy.

\end{abstract} 
\pacs{05.20.-y, 75.10Nr, 87.22.Jb}

Statistical physical modeling of neural networks a\-chieved much
success in the description of neural phenomena, ranging from storage
and retrieval in memory networks to learning and generalization in
feed-forward networks to unsupervised learning \cite{hkp91}.  Whereas
some models for a single neuron are admittedly oversimplified from the
biological viewpoint, when networked they exhibit a variety of neural
functions, performed by living systems and demanded from artificial
designs.  In this Letter we study a single perceptron-type neuron's
memorization ability, crucial for the understanding of networked
systems.  When the number of synaptic couplings of a neuron becomes
large the storage problem can be described via the statistical
mechanical framework introduced by Gardner and Derrida \cite{eg,gd88}.
Since then the neuron is well understood below capacity, the region
beyond it, however, remained the subject of continuous research and
debate \cite{gg90,bou94,mez93,et93,wes96,ws96}.  We claim that the
framework presented here carries the exact statistical mechanical
solution, which we illustrate on a partly analytically treatable
limiting case.  Networks beyond saturation are long known to have
complex features; here we show that even a single neuron can exhibit
extreme complexity.

We consider the McCulloch-Pitts model neuron \cite{hkp91}, \be
\label{defneu} \xi = \text{sign}(h),\quad h = N^{-1/2}
\sum\nolimits_{i=1}^{N} J_i S_i, \ee where $\bf J$ is the vector of
synaptic couplings, $\bf S$ the input and $\xi$ the response.  The
normalization was chosen so that $h$ is typically of $O(1)$ when
$N\rightarrow \infty$.  Patterns to be stored are prescribed as pairs
$\{ {\bf S}^\mu ,\xi ^\mu\}_{\mu =1}^{M}$ such that the neuron is
required to generate ${\xi^\mu}$ in response to ${\bf S}^\mu$.  Given
the ensemble of patterns, the local stability parameter $\Delta^\mu =
h^\mu \xi^\mu$ obeys some distribution $\rho(\Delta)$ (see
\cite{gg90}).  The $\mu$-th pattern is stored by the neuron if the
actual response signal from Eq.\ (\ref{defneu}) equals the desired
output $\xi^\mu$\ie $\Delta^\mu >0$.  The number of patterns $M$ is
generically of order $N$, so $\alpha = M/N$ is an intensive parameter.
For the sake of simplicity, we generate the $S_i^\mu$-s independently
from a normal distribution, consider $\xi^\mu = \pm 1$ equally likely,
and choose the spherical prior constraint $\left|{\bf J}\right| =
\sqrt{N}$. The cost function to be minimized\ie the Hamiltonian, is
the sum of errors committed on the patterns.  The error on the
$\mu$-th pattern is measured by a potential $V(\Delta^\mu)$, taken
here to be zero for arguments larger than a given $\kappa$ and
decreasing elsewhere \cite{gg90}.  Storage as defined above
corresponds to $\kappa = 0$, while a $\kappa >0$ means a stricter
requirement on the local stability $\Delta$ and ensures a finite basin
of attraction for a memorized pattern during retrieval.  The
Hamiltonian defines through gradient descent a dynamics in coupling
space.  Specifically, $V(y)=(\kappa -y )^b\,\theta (\kappa -y)$
corresponds to the perceptron and adatron rules for $b=1,2$,
respectively.  There is no such dynamics in the case $b=0$, but
because of its prominent static meaning -- the Hamiltonian counts the
incorrectly stored patterns -- we will consider that in concrete
calculations.

The Hamiltonian introduced above gives rise to a statistical mechanical
system \cite{eg} resembling models of spin glasses with infinite-range
interactions \cite{sgrev}.  The microstates are configurations of
synaptic couplings, quenched disorder is due to the randomly generated
patterns, and the temperature $T=\beta^{-1}$ represents the tolerance
to error of storage.  The partition function is \be\label{partfunc} Z
= \int_{-\infty}^\infty\!  \!  \! d^N\!  J\ \delta(\sqrt{N} -
\left|{\bf J}\right|) \ \exp \left(-\beta \sum_{\mu =1}^M
V(\Delta^\mu)\right). \ee For large $N$ the replica method
\cite{sgrev} yields the mean free energy per coupling
\cite{eg,gg90,mez93} \be f = - \frac{\left<\ln Z\right>}{N\beta} =
\lim_{n\rightarrow 0} \frac{1-\left<Z^n\right>}{nN\beta} =
\lim_{n\rightarrow 0} \frac{1}{n} \min_{\mbox{\scriptsize{Q}}} f({\sf
Q}) \label{fe}, \ee where $\left< ~ \right>$ stands for the average
over patterns and \bml\label{fe1} \bea f({\sf Q}) & = & f_s({\sf Q})
+ \alpha\, f_e({\sf Q}) , \label{fe1a} \\ f_s({\sf Q}) & = & -
(2\beta)^{-1} \ln\text{det}{\bf\sf Q}, \label{fe1b} \\ f_e({\sf Q}) &
= & - \beta^{-1}\ln \int\!\!\!\int_{-\infty}^\infty\! \!  \! d^n\! x \
d^n\! y\, (2\pi)^{-n} \nn \\& & \times \exp\left( -\beta
\sum\nolimits_{a=1}^n V(y_a) + i{\bf xy} - \case{1}{2} {\bf x} {\sf
Q}{\bf x}\right) \label{fe1c}. \eea \eml The $n \times n$ matrix $\sf
Q$ is symmetric and positive semidefinite, with elements $q_{aa}=1$
and $-1 \leq q_{ab}\leq 1$.  The entropic term $f_s$ is specific to
the spherical model, while the energy-term $f_e$ is independent of the
prior constraint on the synapses.  The mean error per pattern is
%corresponds to the energy and is 
\be \varepsilon =
\frac{1}{\alpha}\frac{\partial\beta f}{\partial\beta}
=\int_{-\infty}^\infty\! \! \! d\Delta \ \rho(\Delta)\, V(\Delta)
\label{ener} 
\ee 
while the entropy per synapsis 
\be s = \beta(\alpha\varepsilon -
f) \label{entr} \ee 
has the usual thermodynamic meaning in coupling
space.

The extremization problem (\ref{fe},\ref{fe1}) was first solved with
the assumption of replica symmetry (RS) \cite{eg,gd88}.  Beyond
capacity at zero temperature, however, Bouten \cite{bou94} showed by
rectifying \cite{eg,gd88} that whenever the local stability
distribution function $\rho(\Delta)$ exhibits a gap, there is an
eigenvalue in negative infinity of the Hessian $\partial ^2f({\sf
Q})/\partial q_{ab}\partial q_{cd}$ at the RS solution, so this is not
a minimum in (\ref{fe}).  Such is the case for the potential
$V(y)=\theta(y-\kappa)$. The one step replica symmetry breaking
($1$-RSB) ansatz was considered for $T = 0$, yielding a $\rho(\Delta)$
different from the RS result, and, as demanded from an improved
solution, a larger energy \cite{mez93,et93,wes96}. In the ground state
beyond capacity, where all $q_{ab}\rightarrow 1$, an eigenvalue of
negative infinity has been found recently for any $R$-step RSB
($R$-RSB), and for illustration the $2$-RSB solution computed
\cite{ws96}.  The results show a slight improvement over $1$-RSB in
the energy and a significant difference in the scaled elements of $\sf
Q$, but also the $2$-RSB ground state turned out to be unstable. Ref.\
\cite{ws96} in fact implied that a gap in $\rho(\Delta)$ at $T=0$
means the instability of all $R$-RSB solutions with $R$ finite.

In order to treat the storage problem of the neuron we technically
generalize Parisi's method for the Sherrington{\-}-Kirkpatrick (SK)
model of spin glasses (see \cite{sgrev}).  By Parisi's choice of $\sf
Q$ and his continuation rule in the $n\rightarrow 0$ limit, the SK
free energy was expressed in terms of an order parameter function.  An
elegant and useful re-formulation was due to \cite{sd84}, whose free
energy functional for the SK problem incorporated both Parisi's and
Sompolinsky's partial differential equations (PPDE and SPDE, resp.).
Its analog was used for the Little-Hopfield (LH) memory network in
\cite{tok94}.  For the neuron, we adopt Parisi's form for $\sf Q$,
momentarily as an ansatz, but thermodynamical stability analysis
reported about later amounts to its consistency check.  Our
calculations show that despite the significant differences between the
SK and the neuron Hamiltonians and those between the 'hard' terms in
the replica free energies, the variational free energies are
remarkably similar.  We obtain \cite{auth}
\vspace{-\baselineskip} \bml\label{fef} \bea f & = & {\ba{c}\mbox{
\footnotesize ~} \\ \mbox{max}\\ \mbox{\footnotesize{\em x(q)}}\ea}
{\ba{c}\mbox{ \footnotesize ~} \\ \mbox{extr}\\
\mbox{\footnotesize{\em f(q,y),P(q,y)}} \ea}\!\!\!  \left[ f_s +
\alpha (f_e + f_a^{(1)} + f_a^{(2)})\right] , \label{fefa} \\ f_s & =
&-(2\beta)^{-1} \int_0^1\!  dq\, \left[ D(q)^{-1} - (1-q)^{-1}
\right], \label{fefb} \\ f_e & = & f(0,0), \label{fefc} \\ f_a^{(1)} &
= & \int_0^1\!\!  dq\ \int_{-\infty}^\infty\!  \!  \!  dy \, P(q,y)
\nn \\ & & \times \left[ \dot{ f} (q,y) +\case{1}{2} f^{\pp} (q,y) -
\case{1}{2}\beta x(q) f^\prime (q,y)^2 \right], \label{fefd} \\
f_a^{(2)} & = &\int_{-\infty}^\infty\!  \!  \!  dy\, P(1,y)\, \left[
V(y)-f(1,y) \right].  \label{fefe} \eea \eml The minimization in
(\ref{fe}) turned to maximization due to its interchange with the
$n\to 0$ limit \cite{sgrev}. Here and later $\dot h = \partial
h/\partial q$ and $h^\prime = \partial h/\partial y$.  The $x(q)$ is
the inverse of Parisi's order parameter function\ie it gives the
probability that the overlap of the synaptic vectors from two replicas
is smaller than $q$, and $D(q)=\int_q^1\! d\bar{q}\, x(\bar{q})$ is
the continuation of the spectrum of the matrix $\sf Q$ for $n \to 0$.
The range $1\ge q\ge 0$ is now included in the ansatz, that should be
verified later.  The auxiliary functionals $f_a^{(1,2)}$ carry the
Lagrange multiplier field $P(q,y)$ and thus vanish at stationarity.
Variation by $P(q,y)$ makes the field $f(q,y)$ satisfy the PPDE, which
can be read off from (\ref{fefd}), and that by $P(1,y)$ fixes the
initial condition through (\ref{fefe}).  So $f(q,y)$ evolves from
$q=1$ to $q=0$ and its final value gives the energy term in
(\ref{fefc}).  Stationarity in terms of $f(q,y)$ and $f(0,y)$ leads to
the SPDE \be \label{spde} \dot{P}(q,y) = \case{1}{2} P^{\pp}(q,y) +
\beta x(q) \left[ P(q,y)\, f^\prime (q,y) \right]^\prime , \ee
evolving from $P(0,y)=\delta (y)$ until $q=1$.  Comparison with the SK
model \cite{sd84}, its $p$-spin generalization \cite{gar85}, and the
LH network \cite{tok94} shows that the respective PDE-s and $P(0,y)$
are the same, but in our case a general initial condition
$f(1,y)=V(y)$ is taken.  In fact, the 'hard' term of the SK replica
free energy is formally a special case of (\ref{fe1c}) if $V(y)=\ln
2\mbox{cosh}y$.  Variation of (\ref{fefa}) in terms of explicit
occurrences of $x(q)$ yields $(2\beta)^{-1} \int_0^1 \!  dq\,
F(q,[x(\bar{q})])\, \delta x(q)$, where \be F(q,[x(\bar{q})]) =
\int_0^q\!  \frac{d\tilde{q}}{D(\tilde{q})^2} - \gamma
\int_{-\infty}^\infty\!  \!  \!  dy\, P(q,y)\, f^\prime(q,y)^2
\label{vari} \ee is simultaneously a function of $q$ and a functional of
$x(\bar{q})$, with $\gamma = \alpha\beta ^2$.  So wherever $\dot{x}(q)>0$
stationarity requires that $F=0$.  If $x(q)\equiv m$, $0<m<1$, in an
interval $I$ then stationarity in terms of $m$ leads to Maxwell's rule
$\int_I\! dq\, F(q,[x(\bar{q})]) = 0$.  The $R$-RSB ansatz involves a
sequence $q_0^{(R)}<\dots <q_R^{(R)}$ and has $x(q)=\sum_{k=0} ^R
(m_{k+1}^{(R)}-m_k^{(R)})\theta(q-q_k^{(R)})$, with $m_0^{(R)}=0\leq
m_1^{(R)}\leq\dots\leq m_{R+1}^{(R)} = 1$.  It is naturally incorporated
into the above scheme: required is $F = 0$ at each of the points
$q_0^{(R)},\dots ,q_R^{(R)}$ and so is the Maxwell rule in the intervals
between them ({\em cf.}  \cite{cs91} in a special case).  Note that the
free energy can be written in short as max$_{x(q)} [f_s+\alpha f_e]$ with
(\ref{fefb},\ref{fefc}), where $f(q,y)$ satisfies the PPDE with the
initial condition as above; that corresponds to Parisi's original 
formulation.

Thermodynamical stability analysis requires the diagonalization of the
Hessian of $f({\sf Q})$ in Eq.\ (\ref{fe1}).  Based on the general
expression of Ref.\ \cite{tdk94} we calculated a subset of eigenvalues
from the replicon sector of the $R$-RSB, including $\lambda^{(R)}(R) =
\dot{F}(q^{(R)}_R,[x(\bar q)])$ that derives from states in the same
smallest cluster.  The $\lambda^{(R)}(R)$ is typically decisive for
stability \cite{cs91,gar85}, and becomes negative infinity at $T=0$
for any $R$-RSB with finite $R$ if $\rho(\Delta)$ has a gap
\cite{bou94,ws96}.  Concerning the maximizing $x(q)$ of (\ref{fef}),
if $\dot{x}(q)>0$ in an interval $I$ then the continuation of the
aforementioned subset is $\lambda(q) =\dot{F}(q,[x(\bar q)])$, so
$\lambda(q)\equiv 0$ in $I$, thus zero modes are present.  This is a
generic property of a Parisi phase \cite{dd96}.

The distribution of the local stability $\Delta$ is found to be of a
remarkably simple form \cite{auth} \be \label{distr}
\rho(\Delta)=P(1,\Delta).  \ee That sheds light on the physical meaning
of the auxiliary field $P(q,y)$: $y$ is the local stability at an
intermediate generation of the ultrametric tree and $P(q,y)$ its
probability distribution.  The analogy with the local magnetic field in
the SK and LH models \cite{sd84,dl83,tok94} is apparent.

Classic neural modeling focuses on $T=0$. To solve that problem,
however, extensive numerical work may be necessary.  On the other
hand, in the limit $\alpha ,\, T\rightarrow \infty$ while $\gamma$ is
kept finite, we can calculate $x(q)$ wherever it deviates from the
step-like shape, thence other analytic results follow. 
By resolving the PPDE and the SPDE perturbatively we obtain $f(q,y)$
and $P(q,y)$ as functionals of $x(q)$ to $O(\beta^2)$, yielding
explicit functional forms for the free energy (\ref{fef}) as well as
for (\ref{vari}).  Another possibility is first expanding (\ref{fe1})
in $\beta$ and then applying the Parisi ansatz.  Either way we arrive
at \bml\label{htfe} \bea \beta^2 f & = & \phi_0+ \beta
\max_{x(q)}\left[\phi_1\right] + O(\beta^2) \\
\label{htfea} \phi_0 & = & \gamma\sqrt{W(0)}
\label{htfeb} \\ \phi_1 & = & \beta f_s + \beta\gamma f_e^{(1)}
\label{htfec} \\ \beta f_e^{(1)} & = & \case{1}{2} \int_0^1\!\!  dq\,
x(q)\, \dot{W}(q) \label{htfed} \\ W(q) & = &
\int\!\!\!\int_{-\infty}^\infty\!  \!  \!  d^2t\frac{\exp\left(
-\case{1}{2} |{\bf t}|^2 \right)} {2\pi} V({\bf n}_1\!\cdot{\bf t})\,
V({\bf n}_2\!\cdot{\bf t}), \label{htfee} \eea \eml where $|{\bf
n}_{1,2}|=1$ and ${\bf n}_1\!\cdot{\bf n}_2=q$.  The functional
(\ref{htfec}) happens to be equivalent with the free energy in
Nieuwenhuizen's generalization of the spherical SK-type spin glass
model \cite{nie95}.  Formula (\ref{vari}) is in leading order
\be\label{htvari} F(q,[x(\bar{q})]) = \int_0^q\!  d\bar{q}\,
D(\bar{q})^{-2} - \gamma\dot{W}(q), \ee thus for a continuous $x(q)$
with $\dot{x}(q)>0$ one has \be\label{htxq} x(q)=\case{1}{2}
\gamma^{-1/2} \overdots{W}(q)\, \ddot{W}(q)^{-3/2}, \ee {\em cf.}
Eq.\ (9) in \cite{nie95}.  Various trial functions $x(q)$, such as an
$R$-RSB, or, Parisi's ansatz of a continuous order parameter function
between two plateaux (such a classic Parisi phase will be referred to
as SG-I), can be formulated by means of (\ref{htvari}).  We calculated
the full set of replicon eigenvalues of $R$-RSB based on \cite{tdk94}.
With $r=0,\dots ,R-1$ and $k,l=r+1,\dots ,R$ we have \be\label{htev}
\lambda (r;k,l)= D(q_k^{(R)})^{-1} D(q_l^{(R)})^{-1}-\gamma
\ddot{W}(q_r^{(R)}), \ee and $\lambda^{(R)}(R)$ is obtained if
$q_R^{(R)}$ is substituted for all $q$-s in (\ref{htev}).  We studied
the example $V(y)=\theta (\kappa -y)$ when \be\label{htwdot}
\dot{W}(q) = (2\pi)^{-1} (1-q^2)^{-1/2}
\exp\left(\kappa^2/(1+q)\right).  \ee Four distinct phases are found
and depicted on Fig.\ \ref{phasediag}.  
%%%%%%%%%%%%%%%%%%%%%%%%%%%%%%%%%%%%%%%%%%%%%%%%%%%%%%%%%%%
%
%    Fig. phasediag
%
%%%%%%%%%%%%%%%%%%%%%%%%%%%%%%%%%%%%%%%%%%%%%%%%%%%%%%%%%%%
At the boundary of the
transition RS---SG-I, furthermore, at the RS---$1$-RSB line for
$\kappa <\kappa_2$, if the border is approached from the RSB phase,
the $x(q)$ function converges for each $0\leq q\leq 1$ to the RS value
$q^{(0)}$.  Here the 3rd derivative of the mean free energy is
discontinuous.  On the other hand, for $\kappa >\kappa_2$, if the
RS--$1$-RSB line is approached from the RSB side then
$q_0^{(1)}\rightarrow q^{(0)}$ but $q_1^{(1)}\not\to q^{(0)}$.  The
plateau value $m_1^{(1)}\rightarrow 1$ so the limits of $x(q)$ from
the two phases differ at one point $q=1$.  At that transition the 2nd
derivative of the free energy is discontinuous.  This phenomenon is
analogous to the RS---$1$-RSB transition in the random energy model
(see \cite{sgrev}), and similar two types of segments of the
RS---$1$-RSB borderline were identified in the spherical, $p$-spin SK
model by \cite{cs91}.  The RS---SG-I boundary is analogous to the
Parisi transition in the SK model.  We found a fourth phase, where
$x(q)$ is like an SG-I curve joined with a 1-step function.  It is of
the same type as the phase PG II of the Potts spin glass \cite{gks85},
and the low-temperature state of the $p$-spin SK model \cite{gar85},
furthermore, it is analogous to the phase SG-IV of \cite{nie96}. The
borderline $\lambda^{(0)}(0) = 0$ of local stability of the RS
state\ie the de Almeida-{\-}Thouless (AT) curve, coincides with the
border of the RS phase for $\kappa <\kappa_2$ but enters the RSB
phases for larger $\kappa$-s.  However, whenever RS and RSB states
coexist, we find that the RSB state maximizes the free energy
functional (\ref{fef}).  No coexistence between different types of RSB
phases was observed.  One characteristic $x(q)$ function from each
phase is shown on Fig.\ \ref{xqfig}.  
%%%%%%%%%%%%%%%%%%%%%%%%%%%%%%%%%%%%%%%%%%%%%%%%%%%%%%%%%%%
%
%    Fig. xqfig
%
%%%%%%%%%%%%%%%%%%%%%%%%%%%%%%%%%%%%%%%%%%%%%%%%%%%%%%%%%%%
Note that if $x(q)$ has a
curved segment, this is explicitly given by Eqs.\
(\ref{htxq},\ref{htwdot}).  For illustration, thermodynamic quantities
are plotted along the $\kappa =0$ line on Fig.\ \ref{extquantfig}. 
%%%%%%%%%%%%%%%%%%%%%%%%%%%%%%%%%%%%%%%%%%%%%%%%%%%%%%%%%%%
%
%    Fig. extquantfig
%
%%%%%%%%%%%%%%%%%%%%%%%%%%%%%%%%%%%%%%%%%%%%%%%%%%%%%%%%%%%
We expect that for some finite temperatures similar phases exist,
nevertheless, in the ground state the phase diagram simplifies to the
single borderline RS---SG-I\ie the known limit of capacity curve.  The
richness of the neural behavior for $T\rightarrow \infty$ should be
contrasted with the generic RS high-{\-}$T$-{\-}phase in SK-type
disordered magnets.

In conclusion, we have put forth an exact description of storage by a
single neuron in terms of a variational free energy, the solution of
which we demonstrated in the high $T$ limit with the error counting
potential.  Storage beyond capacity with other error measures, learning
and generalization of unlearnable tasks, storage by networked neurons,
and frustrated phases in general, are natural directions for future
investigations.

%\acknowledgments{
\bigskip
Special thank is due to T. Temesv\'ari for his patiently explaining
parts of \cite{tdk94}.  Stimulating discussions with C. De Dominicis,
I. Kondor, and the late A. V\'egs\H o are gratefully acknowledged.
This work was supported by HSRF grant No.\ T017272 and the 
Holderbank foundation (Switzerland).
%}

\bfi
\caption{Phase diagram for the potential $V(y)=\theta(\kappa -y)$ in
the $(\gamma ,\kappa)$ plane for high $T$ by numerical maximization of
Eq.\ (\protect\ref{htfec}). The full lines separate phases with
different types of global maxima. The RS, $1$-RSB, SG-IV, and SG-I
phases are indicated by {\protect\em a, b, c}, and {\protect\em d},
respectively.  The AT curve is the RS phase boundary for $\kappa
<\kappa_2\simeq 2.38$ and to the right of the arrow it analytically
continues in the dashed line.}
\label{phasediag} 
\efi

\bfi \caption{The $x(q)$ function at representative points as marked
 on Fig.\ \protect\ref{phasediag} by crosses. }
\label{xqfig} 
\efi

\bfi
\caption{The entropy $s$ from Eq.\ (\protect\ref{entr}), the free
energy term $\phi_1$ from Eq.\ (\protect\ref{htfec}), and the enlarged
correction $\varepsilon_1 =T(\case{1}{2}-\varepsilon)$ for the
energy (\protect\ref{ener}) in the high $T$ limit.  The RS---SG-I
transition is marked by an arrow. The dashed lines
correspond to the thermodynamically unstable RS state beyond this
transition point.}
\label{extquantfig} 
\efi

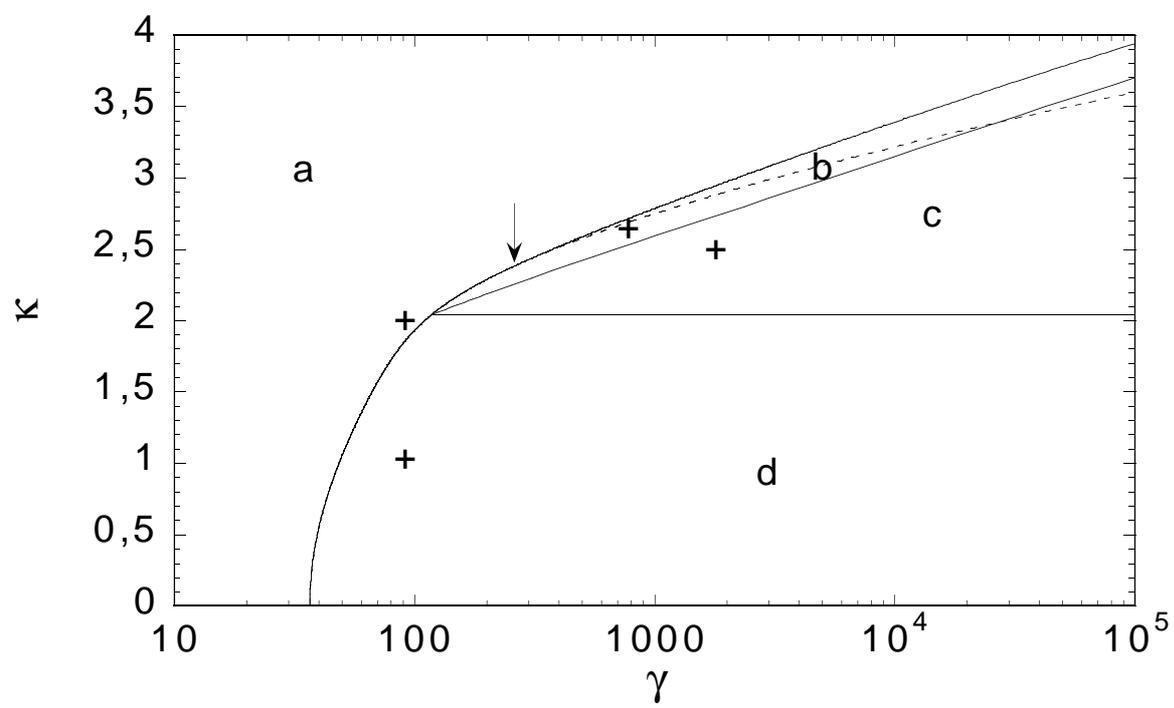
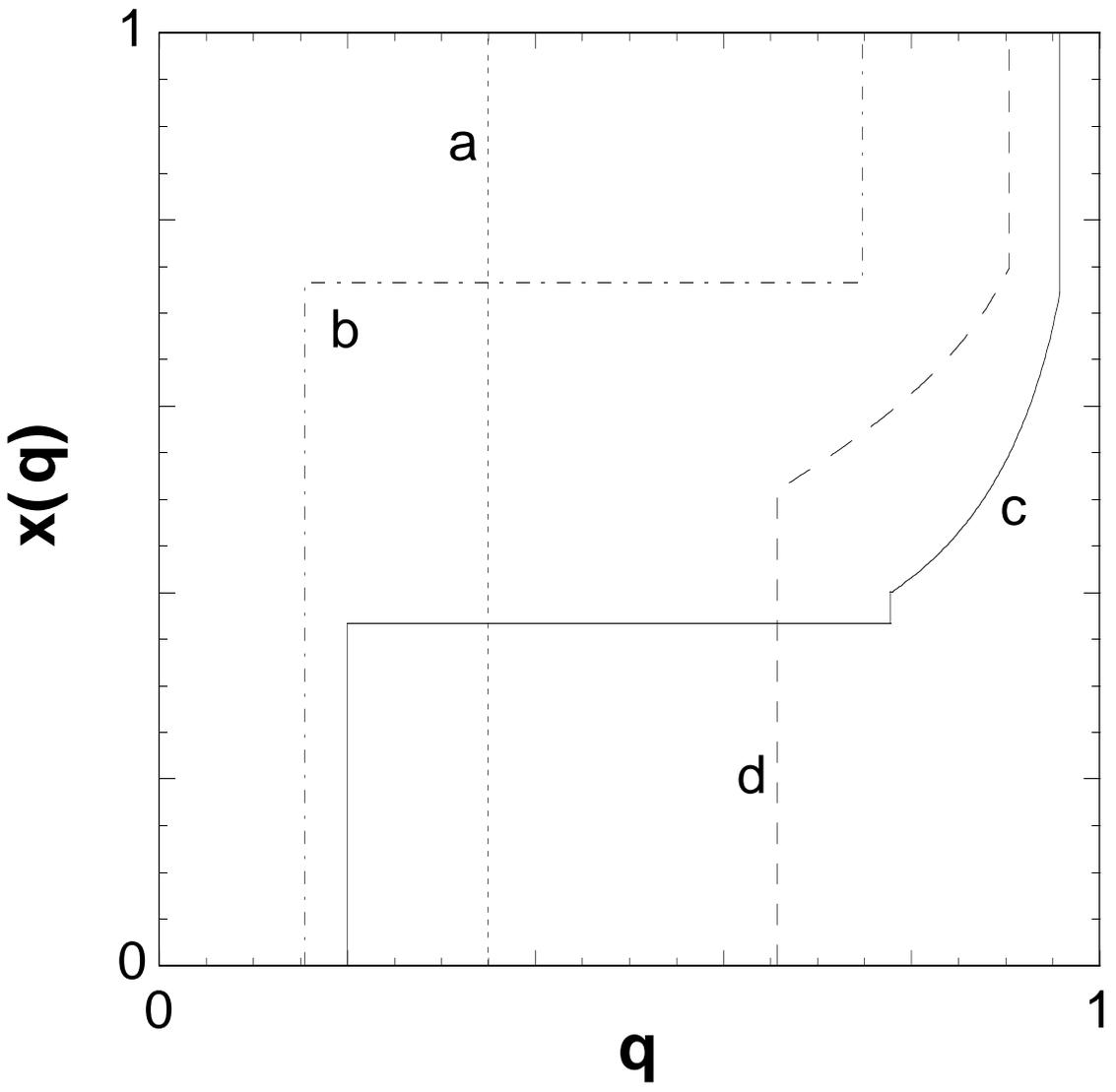
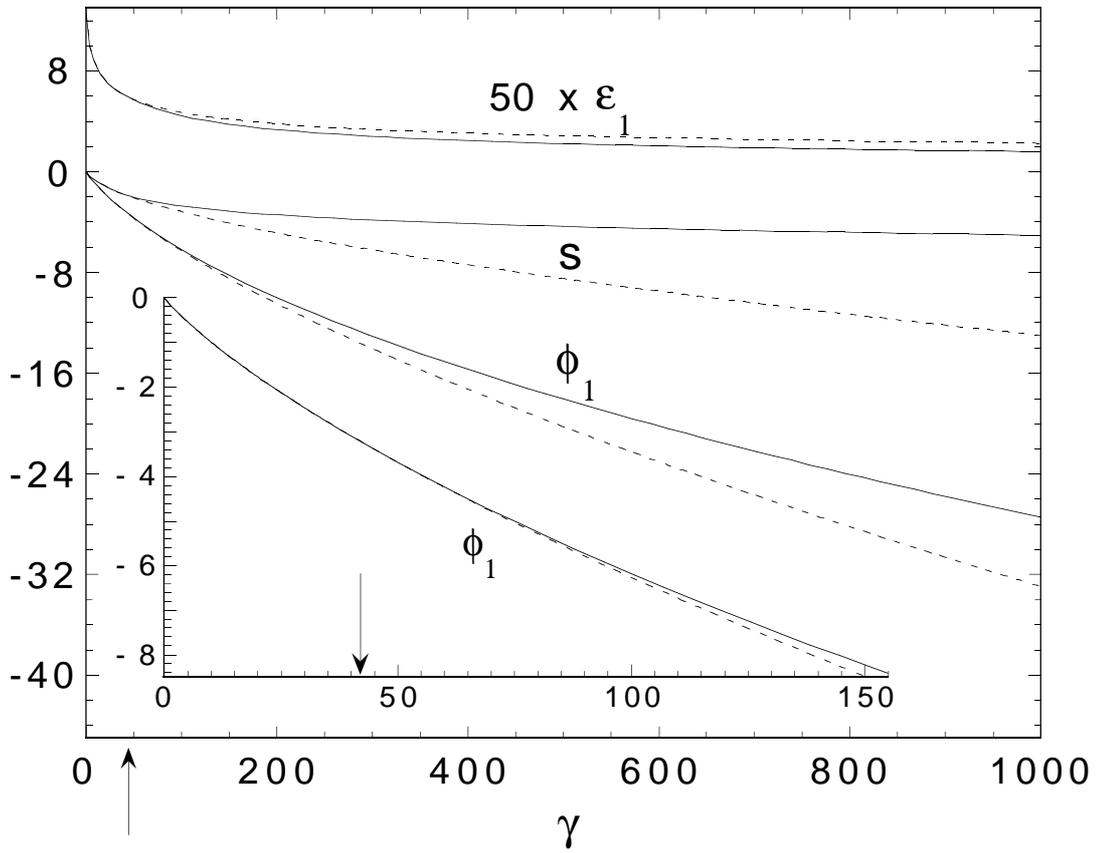
\end{document}